\documentstyle[prd,aps,epsf,epsfig]{revtex} 
\begin{document}
\draft
\twocolumn[\hsize\textwidth\columnwidth\hsize\csname@twocolumnfalse\endcsname
%
\title{
Response in kinetic Ising model to oscillating magnetic fields}
\author
{Kwan-tai Leung and Zolt\'an N\'eda\cite{NZ}}
\address{
Institute of Physics, Academia Sinica,
Taipei, Taiwan 11529, R.O.C.
}

\maketitle
\centerline{\small (Last revised \today)}

\begin{abstract}

Ising models obeying Glauber dynamics in a temporally
oscillating magnetic field are analyzed. 
In the context of stochastic resonance, the response 
in the magnetization is calculated by means of
both a mean-field theory 
with linear-response approximation, and the time-dependent 
Ginzburg-Landau equation. Analytic results
for the temperature and frequency dependent response, including
the resonance temperature,
compare favorably with simulation data.

\end{abstract}

\vspace{2pc}
]

\vspace{1cm}


\section{Introduction}

Ising models with Glauber dynamics in an oscillating magnetic field were
recently considered with Monte Carlo (MC) simulations in  \cite{neda1} and 
\cite{neda2}. The phenomenon of stochastic resonance \cite{stor} was
revealed, exhibiting a characteristic peak in the correlation function 
$C(T)$ 
between the external oscillating magnetic field 
and the magnetization 
$M(t)$ 
versus the temperature $T$ of the system. 
The resonance temperature $T_r$ (the
temperature at which $C(T)$ has a maximum) was systematically computed
as a function of the driving period, 
lattice size
and driving amplitude, 
both for two-dimensional 
(2D) \cite{neda1} and three-dimensional (3D) \cite{neda2} systems. 
The one-dimensional (1D) case was analysed by Brey and Prados 
\cite{brey} in a linear-field approximation.

The present work is a natural continuation of 
those studies, considering 
analytically the 2D and 3D cases. We will present two approaches. 
The mean-field 
theory with linear response approximation will be discussed first. Then
in 2D where the mean-field theory is not as good as in
other dimensions, a more refined time-dependent 
Ginzburg-Landau (TDGL) approach will be presented, 
with significant improvements. 

Recently, kinetic Ising systems in 
oscillating external fields have also been examined 
both experimentally 
and theoretically in \cite{applphys}. 
The focus was on properties below the zero-field critical point,
such as the frequency dependence of the probability 
distributions for the hysteresis-loop area and the residence time.
The latter quantity for small systems in moderately weak 
fields suggests further evidences of stochastic resonance.
Very recently,  finite-size effects versus driving
frequency are analyzed as a dynamical critical phenomena \cite{new}.
In contrast to these works, ours is focused on the temperature
dependence above the zero-field critical point.

\section {Mean-field theory and linear-response approximation}

Our starting point is the master equation for the kinetic Ising model
obeying Glauber dynamics \cite{Glauber}:
\begin{eqnarray}
P(&{\sigma}&;t+1)-P({\sigma};t)= \nonumber\\
&~&\sum_{\sigma'}
       \left[ w({\sigma'}\to {\sigma})P({\sigma'};t)
              -w({\sigma}\to {\sigma'})P({\sigma};t) \right],
\label{master}
\end{eqnarray}
where $P({\sigma};t)$ is the joint probability of finding the spin 
configuration ${\sigma}$ at time $t$, and $w$'s are the transition rates 
between two configurations which differ by one spin flip.
For the heat-bath algorithm, the rate function is chosen as
\[
w({\sigma}\to {\sigma'})=
       { 1 \over 1+e^{-\beta[E(\sigma)-E(\sigma')]} },
\]
with $\beta=1/ T$ 
(hereafter the Boltzmann constant $k\equiv 1$),
and $E(\sigma)$ is the energy of $\sigma$ in a magnetic field $h$:
\begin{equation}
E(\sigma)=-J \sum_{\rm nn} S_i S_j- h(t) \sum_i S_i,
\label{energy}
\end{equation}
where $h(t)=A \sin(\omega t)$ and
$\sum_{\rm nn}$ denotes a summation over nearest neighbors 
in a square or cubic lattice.

Let us denote the configuration $\sigma$ by the values of the spins
$S_1, S_2, ...,S_V$, with system volume given by $V=N^d$.
$d$ is the spatial dimension of the system and $N$ is its linear size.
Since $S_i=\pm 1$, 
it is easy to rewrite (\ref{master}) as
\begin{eqnarray}
\frac{d}{dt} &P&(S_1, S_2,.....S_V; t)=
 -\sum_{j=1}^{V} w_j(S_j) P(S_1, S_2,...,S_V; t)
\nonumber\\
&+&\sum_{j=1}^{V} w_j(-S_j) P(S_1, S_2,...,-S_j,...,S_V;t)  
\label{master2}
\end{eqnarray}
with
\begin{eqnarray}
w_j(S_j) &=& \frac{1}{2}[1-S_j \tanh(E_j/T)],
\label{prob}\nonumber\\
E_j &=& J \sum_{k=1}^{z} S_k + h,
\label{eng}
\end{eqnarray}
where the last sum runs over the $z$ nearest neighbors of the 
spin $S_j$, with $z=2d$.
Multiplying both sides of (\ref{master2}) by $S_l$
and performing an ensemble average (denoted by $<\cdots>$),
after some simple mathematical tricks, 
we get the basic equation for the Glauber dynamics:
\begin{equation}
\frac{d}{dt} <S_l> =-<S_l>+<\tanh(E_l/T)>.
\label{basic}
\end{equation}
Invoking the mean-field approximation, we replace
$E_l$ by $Jz<S>+h$ to get:
\begin{equation}
\frac{d}{dt} <S>=-<S>+\tanh[(h+ T_c^{\rm MF} <S>)/T],
\label{mfe}
\end{equation}
where $T_c^{\rm MF}=Jz$ is the mean-field critical temperature.
In the absence of $h$, the magnetization is given by the 
stationary solution of the well-known equation:
\begin{equation}
<S>_0=\tanh[ T_c^{\rm MF} <S>_0/T].
\label{stacionary}
\end{equation}
For small $h(t)$,
we may use the linear-response theory in (\ref{mfe})
by first writing 
$<S>(t)=<S>_0+\Delta S(t)$
and considering the $h/T\ll 1$ and $\Delta S/T\ll 1$ limits. 
Performing the Taylor expansion and keeping 
only the first-order terms, equation (\ref{mfe}) becomes
\begin{equation}
\frac{d}{dt} \Delta S= - \frac{\Delta S}{\tau_{\rm MF}} + \frac{A}{T}
(1-<S>_0^2) \sin(\omega t),
\label{calc}
\end{equation}
where
\begin{equation}
\tau_{\rm MF}=\frac{1}{1-   \frac{ T_c^{\rm MF} }{T}   (1-<S>_0^2)}
\label{tau}
\end{equation}
is the relaxation time.
The solution can be found easily:
\begin{equation}
\Delta S(t)= \Delta S_0 \sin(\omega t-\theta_{\rm MF}),
\label{lra2}
\end{equation}
with the phase shift and amplitude given by 
\begin{eqnarray}
\theta_{\rm MF} &=&\arctan(\omega \tau_{\rm MF})
\\
\label{phase}
\Delta S_0&=&\frac{A}{T}(1-<S>_0^2)
\frac{1}{\sqrt{ {1\over \tau_{\rm MF}^{2} }+\omega^2}}.
\label{response}
\end{eqnarray}
The correlation function between the total magnetization
$M=V <S>$ and the external field $h(t)$ can be computed:
\begin{eqnarray}
C&=& \overline {M(t) h(t)}\equiv(V\omega/2\pi) \int_0^{2\pi/\omega} 
\Delta S(t) h(t) dt
\nonumber\\
\label{corcalc}
&=&\frac{V A^2}{2T} (1-<S>_0^2) 
\frac{\tau_{\rm MF}}{1+\omega^2 \tau_{\rm MF}^2}. 
\label{mfcorel}
\end{eqnarray}
Here the overline denotes a temporal average over a period $P=2\pi/\omega$.
In the $T>T_c^{\rm MF}$ domain, $<S>_0=0$, thus $C$ becomes:
\begin{equation}
C_{T>T_c^{\rm MF}}=\frac{V A^2}{2} \frac{T-T_c^{\rm MF}}
{(T-T_c^{\rm MF})^2+\omega^2T^2}.
\label{mfc_above}
\end{equation}

\section{Time-dependent Ginzburg-Landau approach}

Before comparing (\ref{mfcorel}) to simulations,
we present an alternative, continuum approach to compute $C$.
For an Ising system with non-conservative order parameter 
(model A\cite{hh}),
the time-dependent Ginzburg-Landau (TDGL) equation 
for the local magnetization density $\phi(\vec{r},t)$ 
takes the following form:
\begin{eqnarray}
{\partial\phi\over \partial t} 
&=& -\Gamma {\delta {\cal H}\over \delta \phi}+\zeta,
\label{tdgl}\\
{\cal H} &=& 
\int d\vec{r}\,\left\{ {1\over 2}(\nabla\phi)^2 + {1\over 2}u\phi^2
             + {g\over 4!}\phi^4 \right\},
\label{ham}
\end{eqnarray}
where $\cal H$ is the coarse grained Hamiltonian.
For our present purpose, the
white noise $\zeta(\vec{r},t)$ which accounts for the effect of
thermal fluctuations is irrelevant.
Conventionally, 
parameters $\Gamma$, $u$ and $g$ in (\ref{ham}) are 
understood to be obtained by coarse graining
the microscopic dynamics (\ref{master}).
For critical properties, the important temperature dependence
in these parameters lies in $u\propto T-T_c^{\rm GL}$, giving rise to the 
spontaneous symmetry breaking below the critical temperature
$T_c^{\rm GL}$.
In order to compare with simulations,
more precise dependences on $T$ are required.
To this end, we outline here a refined mean-field approach 
in the continuum limit.
The same approach has been successfully applied to the two-species
driven diffusive systems\cite{prl94}.
This approximation is expected to be good outside the critical
region. However, this turns out to be not a serious handicap
because the presence of an oscillating field prevents the system
from building up critical correlations.

In a mean-field approximation, the joint probabilities in (\ref{master})
are factorized into singlet probabilities $p({\vec r};t)$ for
finding the spin up at site $\vec r$ at time $t$.
This effectively produces the power series expansion of $\cal H$
in $\phi$.
$1-p$ gives the probability of finding the spin down.
The continuum limit involves an expansion in the derivatives, such as:
\[
p(x\pm 1,y;t)\rightarrow p(x,y;t)
              \pm {\partial p(x,y;t)\over \partial x}
              +{1\over2}{\partial^2p(x,y;t)\over \partial x^2} +\cdots 
\]
By identifying $p$ as $(\phi+1)/2$,
we obtain from (\ref{master}) 
a kinetic equation for $\phi$ after some algebra.
For $h=0$, we find precisely the deterministic part of (\ref{tdgl}) with:
\begin{eqnarray}
\Gamma&=&{1\over 8}  (      -2W_4 + 2W_{-4} -  W_8 + W_{-8}),
\label{Tdep_G}\\
u&=&{1\over 8 \Gamma}( 6W_0+12W_4 - 4W_{-4} + 5W_8 -3W_{-8}),
\label{Tdep_u}\\
g&=&{3\over 2 \Gamma}(-6W_0-4 W_4 + 4W_{-4} + 5W_8 + W_{-8}),
\label{Tdep_g}
\end{eqnarray}
where $W_n\equiv 1/(1+e^{n\beta J})$ contains the desired 
explicit $T$ dependence.
When a small uniform field $h$ is applied,
to $O(h)$ (neglecting a $\phi^5$ term) we have finally 
the deterministic kinetic equation
\begin{equation}
{\partial\phi\over\partial t} 
= -\Gamma \left\{-\nabla^2\phi+u\phi+{g\over 6}\phi^3 -\mu h \right\},
\label{tdglexp}
\end{equation}
where $\mu=\beta (3W_0^2+4W_4 W_{-4} + W_8 W_{-8})/2\Gamma$.
It is useful to note that $\Gamma$, $g$ and $\mu$ in (\ref{tdglexp}) 
are positive definite for all $T$, whereas $u$ has one zero at
$T_c^{\rm GL} \approx 3.0901 J\approx 1.3618 T_c$, where 
$T_c=-2/\ln(\sqrt{2}-1)J\approx2.2692J$ is exact.  This is an improvement
over $T_c^{\rm MF} =4 J$ from the last section. 
Moreover, we reproduce the first few terms
of the high-temperature series expansions of thermodynamic quantities
such as the susceptibility and the relaxation time.
In the $\beta\to 0$ limit, we recover the mean-field results
of the last section:
$u \approx 1/\beta J -4$,
$\Gamma\approx\beta J$, $g\approx 48(\beta J)^2$, and $\mu\approx 1/J$.

For small $h$ and 
$T>T_c^{\rm GL}$, 
the nonlinear term $g\phi^3$ in (\ref{tdglexp}) 
is negligible. The total magnetization
$M(t)=\int d{\vec r}\,\phi(\vec{r},t)=\tilde \phi({\vec q=0},t)$ 
in response to an external 
field can then be computed easily, where
$\tilde \phi$ denotes the spatial Fourier transform of $\phi$.
It satisfies $\partial M/\partial t=-\Gamma u M 
+ \Gamma \mu \tilde h({\vec q=0},t)$.
We readily find
\begin{equation}
M(t)= {V \mu A\Gamma\over \sqrt{(\Gamma u)^2+ \omega^2} }
\sin(\omega t-\theta_{\rm GL}),
\label{M}
\end{equation}
where the phase shift is $\theta_{\rm GL}=\arctan(\omega/\Gamma u)$.
The correlation function with $h$ is then given by
\begin{equation}
C_{T>T_c^{\rm GL}}=
{V A^2 \Gamma^2 \mu u\over 2[(\Gamma u)^2+ \omega^2] }.
\label{corr_above}
\end{equation}
Note that this coincides with the mean-field result 
(\ref{mfc_above}) in the high-temperature limit.

For $T<T_c^{\rm GL}$, the term proportional to $g$ 
is needed to break the symmetry, leading
to the spontaneous magnetization $m=\sqrt{-6u/g}$.
Linearizing about $m$, we find precisely the same 
form of $C$ as $T>T_c^{\rm GL}$ except 
that $u$ is replaced by $-2u$ in (\ref{corr_above}).

\section{Discussion and comparison with simulations}

>From the simulation data in \cite{neda1} and \cite{neda2},
we learn that the
system has a maximum response to external driving at a definite temperature
$T_r$ which depends on the driving frequency. 
Hence $T_r$ can be designated as the {\em resonance temperature}. 
>From the analytically determined correlation functions 
in (\ref{mfc_above}) and (\ref{corr_above}), we find 
two peaks in $C$ at above and below the respective $T_c$,
and also $C(T_c)=0$,  as shown in Fig.~1. 
This double-peak structure in $C$ is consistent with simulations
for larger lattice sizes 
(up to $N=200$ for 2D and $N=40$ for 3D) and with smaller 
steps in $T$ than reported in \cite{neda1} and \cite{neda2}. 
The reason for missing the peak below $T_c$ 
in our earlier simulations is due to the use of small lattice sizes.
Note that the peak below $T_c$ is much smaller than the one above 
and its position is less sensitive to the driving period. 
The reason for the overestimated theoretical values of the peaks
below $T_c$ may be accounted for by the frustration of the system to order
in the presence of $h(t)$; such frustration steming from metastability
has not been taken into account explicitly in our theories below $T_c$
when we seek the symmetry-breaking solutions.

We believe that this also explains the discrepancy
at $T_c$, where simulations show a small but finite $C(T)$.
Finite-size effects are not of great concern here because,
as mentioned above, the correlation length even at $T_c$
is truncated by $h$.
In simulations, we have checked the convergence in $C(T)$
for $N\geq 50$ in 2D.

Focusing on $T>T_c$ from now on,
the TDGL predictions for $C(T)$
are more accurate than 
those of the mean-field theory in general.
They both converge to the simulations
in the tails at $T\gg T_c$ (see Fig.~1).
In 3D the mean-field theory is already acceptable.

Turning our attention to the amplitude dependence,
replotting the simulation data from \cite{neda1} and \cite{neda2} suggests
that the height of the peak $C(T_r)\propto A^2$, 
in agreement with (\ref{mfc_above}) and (\ref{corr_above}).
For not too large frequencies and small $A$,
the theoretical proportionality constant agrees well with simulations.
For example, 
the slope of $C(T_r)/(V T_c)$ versus 
$A^2/T_c^2$ for $P=50$ in 2D 
gives $0.92$ from simulations
\cite{neda1}, $0.96$ from TDGL and $0.99$ from mean-field
approach. In 3D the same slope is $0.88$ from simulations 
\cite{neda2}, and $1.29$ from mean-field approach
(In 3D the comparison are worse
because $T_r$ is much closer now to $T_c$.)
This proportionality is a manifestation of the linear response
of the system to $h$, which breaks down at large enough amplitudes.
Our new simulations show that this happens for
$A/T_c>0.15$ in 2D for $P=40$.

A quantity of significant interest is the resonance temperature $T_r(P)$.
It can be determined analytically from (\ref{mfc_above})
\begin{equation}
T_r^{\rm MF}=T_c^{\rm MF}\left(1+\sqrt{1-\frac{1}{\omega^2+1}}\right),
\label{res_temp}
\end{equation}
and numerically from (\ref{corr_above}) for $T_r^{\rm GL}$. 
These together with simulation results
are presented in Fig.~2. The agreements are reasonable.
As expected the 
mean-field approximation is quite good in 3D 
but in 2D the TDGL approximation is better.

The results in Fig.~2 confirm the earlier observation in \cite{neda1} and
\cite{neda2} that for 
$P\rightarrow \infty$ we get $T_r\rightarrow T_c$. This result is also
consistent with the one obtained by Brey and Prados \cite{brey} in 1D
where the above limit becomes $T_r \rightarrow T_c=0$. In the
opposite limit $P\rightarrow 1$ (in unit of Monte Carlo steps
$P\ge 1$) both the theory in 1D \cite{brey} and our approximations 
in 2D and 3D suggest $T_r \rightarrow {\rm const}$. 
Unfortunately, in \cite{neda1} 
and \cite{neda2} the wrong conclusion 
$T_r\rightarrow \infty$ was drawn in this limit.
Similarly, the position of the peak below $T_c$ also converges to $T_c$
in the $P\rightarrow \infty$ limit.

In passing, 
we also derive \cite{KTL-ZN} the relationship between 
the correlation function and the hysteresis-loop area $\cal A$:
\begin{equation}
{\cal A}= 2 \pi C \mid \tan\theta \mid
\label{hyst}
\end{equation}
where $\theta$ is the phase shift between $h$ and $M$.
This then relates our results of $C$ to that of $\cal A$ 
as observed in \cite{applphys}.

\section{Conclusions}

Using mean-field with linear-response and TDGL approximations, the
characteristics of the resonance peaks observed in kinetic Ising models
in oscillating magnetic fields \cite{neda1,neda2}
are reproduced.
New simulations improve
earlier results by confirming the analytically predicted double peaks. 
Focusing mostly on the behavior above $T_c$ (where our approaches work 
better), we determine the dependence of the resonance temperature 
as a function of driving frequency and amplitude. 
We confirm the already predicted result
in \cite{neda1,neda2} that $T_r\rightarrow T_c$ for the limit
of practically interesting driving frequencies ($P\rightarrow \infty$),
and corrected the wrong extrapolation in
the opposite limit $P\rightarrow 1$.
We introduce a refined TDGL approach which improves significantly 
the mean-field results in 2D,
but in 3D the mean-field approximation is already acceptable.
We have thus demonstrated that the stochastic resonance
in kinetic Ising models above $T_c$ can be understood 
by means of rather simple theoretical approaches
for small driving amplitudes.

\section{Acknowledgments}

We are grateful to NSC of ROC for their support through the grant
NSC87-2112-M-001-006.



\onecolumn


\begin{figure}[htp]
\epsfig{figure=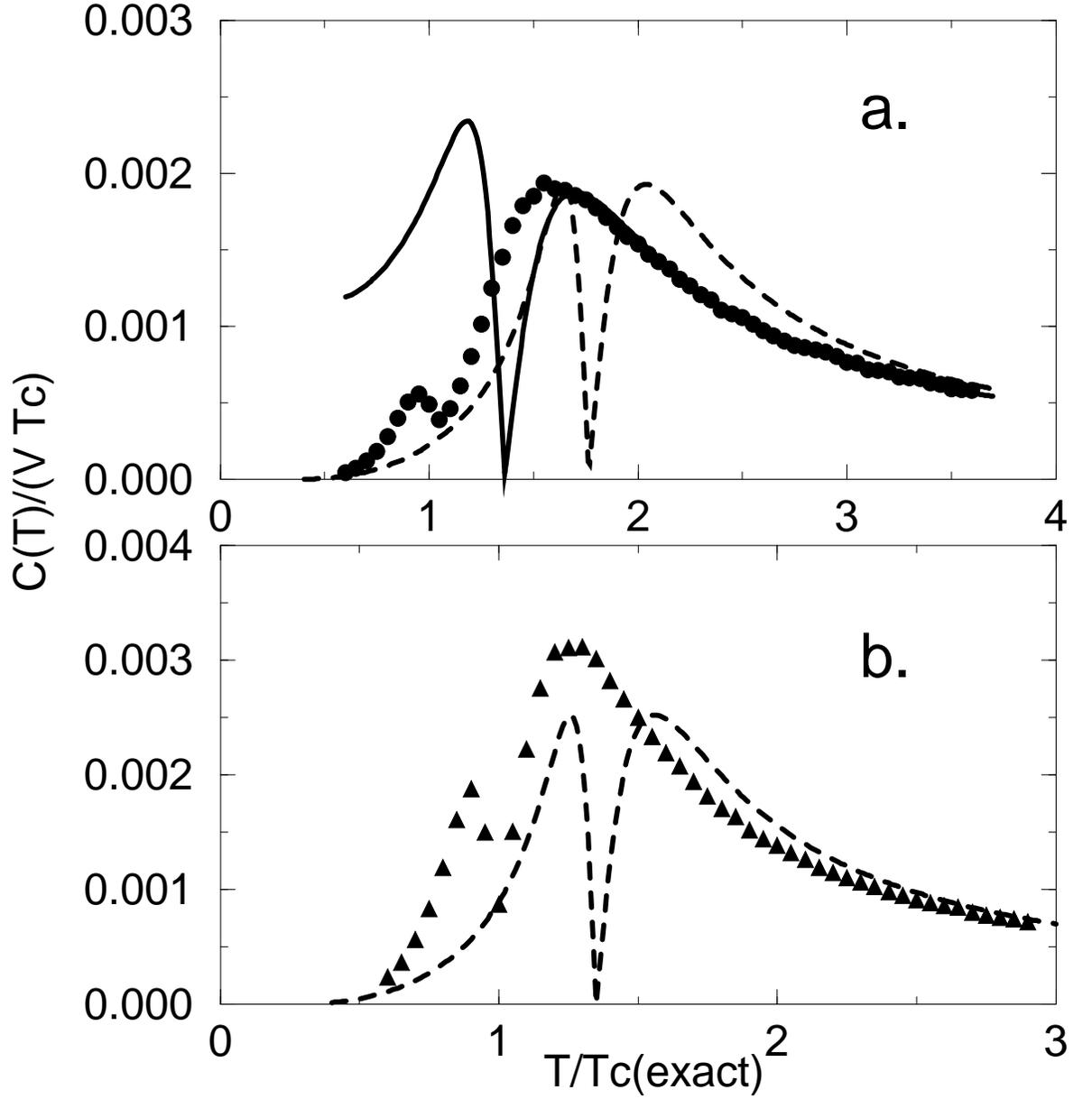,height=7.0in,width=9.0in,angle=-0}
\caption{$C(T)/(V T_c)$ versus temperature for $P=40$ and $A=0.05T_c$
for 2D in (a) and 3D in (b).
Dots are MC simulation results in 2D ($N=200$), triangles
are MC simulations in 3D ($N=40$), 
continuous line is from TDGL approximation  
and the dashed line is the mean-field result. 
The higher peak for TDGL than mean-field theory below 
$T_c$ is due to our dropping the $\phi^5$ term in 
(20).}
\label{fig1}
\end{figure}


\begin{figure}[htp]
\epsfig{figure=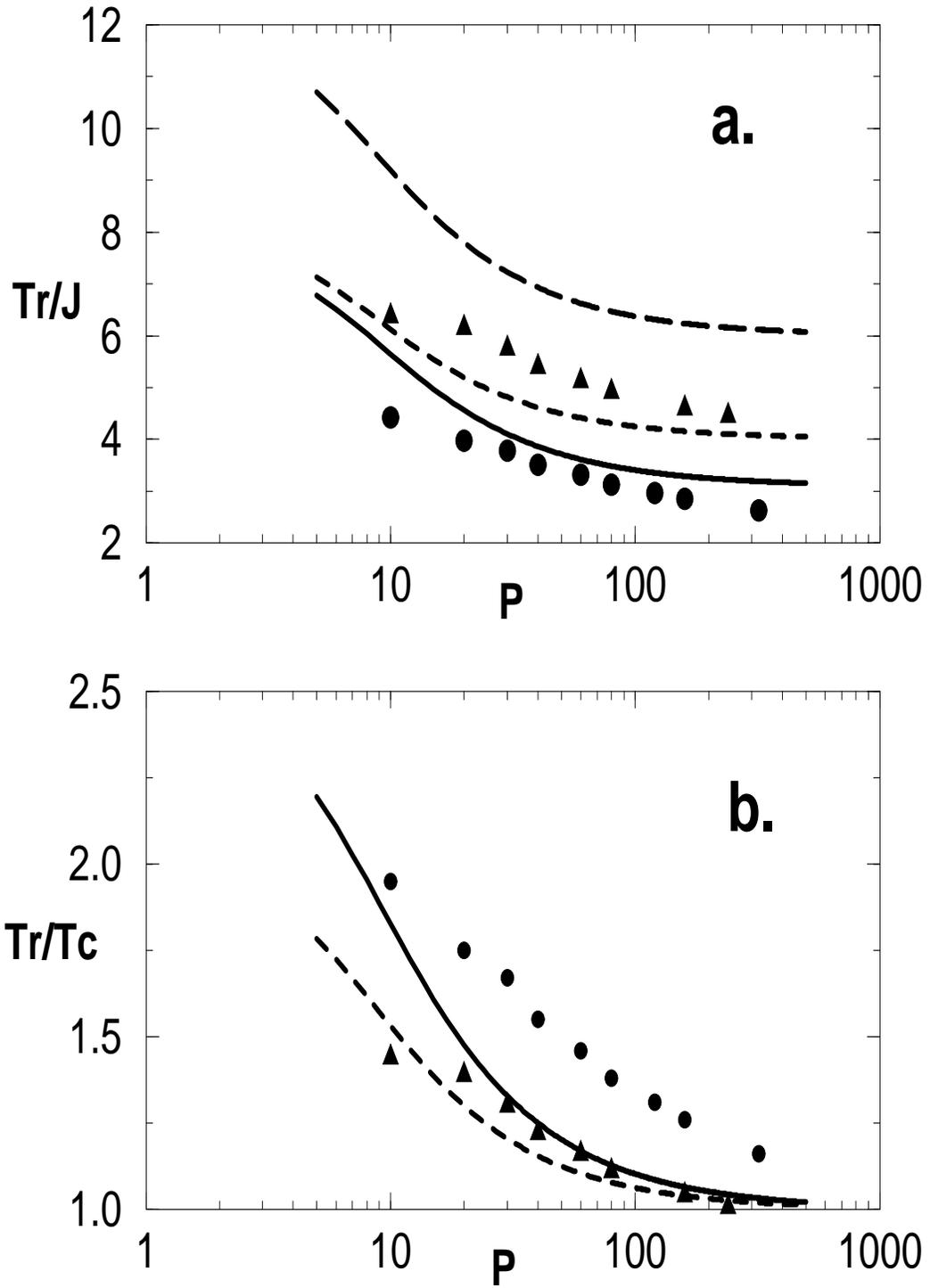,height=8in,width=8.0in,angle=-0}
\caption{Resonance temperature above $T_c$ versus driving 
period $P$ for $A=0.05$, on absolute
scale $T_r/J$ in (a) and on relative scale $T_r/T_c$ in (b).
The long-dashed and short-dashed lines in (a) are the mean-field 
results for 3D and 2D respectively, in the
rest the symbols mean the same as in Fig.~1.
}
\label{fig2}
\end{figure}

\end{document}